# Encapsulated graphene based Hall sensors on foil with increased sensitivity

**Zhenxing Wang**[*,1], **Luca Banszerus**[2], **Martin Otto**[1], **Kenji Watanabe**[4], **Takashi Taniguchi**[4], **Christoph Stampfer**[2,3], **Daniel Neumaier**[1]

[1] Advanced Microelectronic Center Aachen (AMICA), AMO GmbH, Otto-Blumenthal-Straße 25, 52074 Aachen, Germany
[2] JARA-FIT and 2nd Institute of Physics, RWTH Aachen University, 52074 Aachen, Germany
[3] Peter Grünberg Institute (PGI-9), Forschungszentrum Jülich, 52425 Jülich, Germany
[4] National Institute for Materials Science, 1-1 Namiki, Tsukuba 305-0044, Japan

[*] Corresponding author: e-mail wang@amo.de, Phone: +49 241 8867 210, Fax: +49 241 8867 571

The encapsulation of graphene based Hall sensors on foil is shown to be an effective method for improving the performance in terms of higher sensitivity for magnetic field detection. Two types of encapsulation were investigated: a simple encapsulation of graphene with polymethyl methacrylate (PMMA) as a proof of concept and an encapsulation with mechanically exfoliated hexagonal boron nitride (hBN). The Hall sensor with PMMA encapsulation already shows higher sensitivity compared to the one without encapsulation. However, the Hall sensor with graphene encapsulated between two stacks of hBN shows a current and a voltage normalized sensitivity of up to 2270 V/AT and 0.68 V/VT respectively, which are the highest reported sensitivity values for Hall sensors on foil so far.

**1 Introduction** Hall sensors are broadly used in modern automatic systems for switching and position detection [1-3]. The Hall sensors based on graphene show higher sensitivity compared to the most widely used commercial Hall sensors, which are fabricated based on silicon technology [4-6]. In combination with the high flexibility of graphene, corresponding applications in automobile and communication systems are expected [6,7]. The sensitivity of Hall sensors for magnetic field detection depends on the sheet carrier density $n$ and the charge carrier mobility $\mu$, and thus lower $n$ or higher $\mu$ results in higher sensitivity [1,2]. If a graphene based Hall sensor is characterized in ambient, the exposure to air usually introduces p-type doping to graphene due to the physical adsorption of oxygen or water molecules, leading to a higher carrier density and hence a lower sensitivity [8,9]. For graphene based Hall sensors on rigid substrates, e.g. silicon with silicon oxide, an electrostatic doping from the global back gate can be utilized to compensate the doping from the air resulting in a lower doping level and thus a higher sensitivity. However, for flexible graphene based Hall sensors on foil, e.g. Kapton, due to the lack of a back gate, an encapsulation layer is expected to be a better choice for effectively reducing the carrier density and hence increasing the sensitivity of the Hall sensor. Meanwhile, the encapsulation isolates the active material graphene from the environment and thereby enables a stable operation for an actual application. Here, we show that the encapsulation of the graphene can improve the performance of Hall sensors on foil in terms of the increasing sensitivity for graphene produced by chemical vapour deposition (CVD) growth. This work provides an effective performance optimization strategy for graphene based Hall sensors on foil, making this technology more feasible for further applications.

**2 Experimental** The Hall sensors in this work were all fabricated on 50 $\mu$m thick Kapton foil with photolithography as the patterning technology. Prior to the fabrication, the Kapton foil was fixed on a silicon chip with polydimethylsiloxane (PDMS) coating for better handling of the foil.

Afterwards the Kapton was coated with 600 nm SU-8 photoresist to create a smooth surface [7]. Two encapsulation methods were investigated: encapsulation of graphene by polymethyl methacrylate (PMMA) and hexagonal boron nitride (hBN). For the sample with PMMA encapsulation, commercially available large scale monolayer graphene grown by CVD was used (Graphenea SE). It was transferred to the chip using a PMMA-assisted method [10], and the graphene was then patterned with oxygen plasma, followed by metallization of 50 nm nickel to contact graphene. Before the encapsulation, the samples were pre-baked at 100ºC for 10 minutes. The encapsulation was carried out by covering the active area with a drop of PMMA (950K), and following baking from room temperature up to 180 ºC for 10 minutes to remove the solvent. For the sample with hBN encapsulation, large and isolated single crystal domains were grown by CVD on copper foil in house and micromechanical exfoliated multilayer hBN was used as the encapsulation layer (details are described in Ref. [11]). The hBN-graphene-hBN sandwich structure was directly used for the sensor fabrication. The complete stack was patterned by $SF_6$/Ar plasma etching with aluminium as hard mask, which was then stripped with wet chemical etching [5]. Afterwards 50 nm nickel was deposited to contact graphene from the side. All the measurements were carried out in ambient atmosphere at room temperature with a probe station and a semiconductor parameter analyser (HP 4156B). The magnetic field was applied by an electromagnet whose electric field was calibrated in advance with a commercial Hall sensor (Allegro Microsystems A1324).

**3 Hall sensors encapsulated with PMMA** A symmetric cross geometry is utilized for the Hall sensor used for PMMA encapsulation, and the as fabricated sample before encapsulation is shown in Fig. 1a, with a detailed optical image shown in Fig. 1b. The distance between the two opposite electrodes $L$ is 500 $\mu$m and the width of the graphene strip $W$ is 200 $\mu$m. The Hall measurements were carried out before and after the PMMA encapsulation. During the Hall measurement, a constant voltage supply was applied across one pair of opposite electrodes and set to 0.1 V. The potential difference under an applied specific magnetic field is measured over the other pair of opposite electrodes, which is known as the Hall voltage. The Hall measurement was carried out with magnetic fields of 4.5 mT, 8.9 mT, 13.4 mT, -13.4 mT, -8.9 mT, -4.5 mT in sequence, and the magnetic field was set to zero between different applied magnetic field values. The response of the Hall voltage to the external magnetic field is plotted in Fig. 1c. As clearly shown in the figure, the offset of the Hall sensor, i.e. the measured Hall voltage while the magnetic field is zero due to the geometrical imbalance, is stable without any obvious drifting throughout the measurement, which is crucial for practical applications. The current flowing through graphene is 83 $\mu$A dur-ing the measurement. The Hall voltages for different magnetic fields are averaged over time and the relationship with respect to the external field is plotted in Fig. 1d, in which a clear linear relationship is shown. The relative errors for the Hall voltage measurement are also shown for different magnetic fields, and all the errors are less than 0.3%. The linear fitting to the data points shows a slope of 0.0043 V/T with a fitting error of $2.4 \times 10^{-5}$ V/T, corresponding to a voltage and a current normalized sensitivity of $S_V$ = 0.043 V/VT (with an error of $2.4 \times 10^{-4}$ V/VT from the fitting) and $S_I$ = 52 V/AT (with an error of 0.29 V/AT from the fitting) respectively, based on the bias voltage of 0.1 V and the current of 83 $\mu$A. These values are similar to that from Hall sensor devices on foil with smaller size, which demonstrates the uniformity of the large scale single-layer graphene film [7].

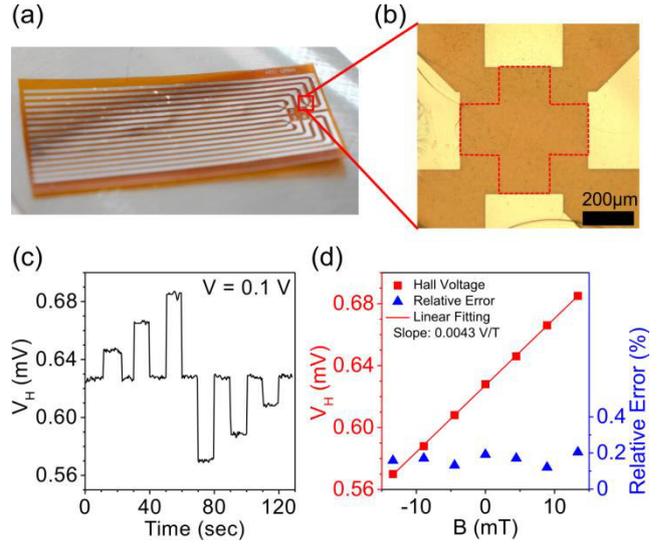

**Figure 1** The Hall sensor based on large area CVD graphene on foil. (a) The as fabricated sample containing the Hall sensor. (b) The optical image of the active area of one Hall sensor with a scale bar of 200 $\mu$m. The graphene area is marked with dashed lines. The length and width of the graphene strip are 500 $\mu$m and 200 $\mu$m respectively. (c) The Hall voltage response to the applied magnetic field, with a sequence of 4.5 mT, 8.9 mT, 13.4 mT, -13.4 mT, -8.9 mT, -4.5 mT. Throughout the measurement, the bias is set to 0.1 V, and a constant current of 83 $\mu$A is measured. (d) The linear relationship between the Hall voltage and magnetic field, with a slope of 0.0043 V/T from the linear fitting to the data points. The calculated voltage and current normalized sensitivity are 0.043 V/VT and 52 V/AT respectively. The relative errors for the Hall voltage measurement are also shown.

After the encapsulation with PMMA, the obtained voltage and current normalized sensitivity are 0.064 V/VT and 121 V/AT respectively. The sensitivity $S_V$ and $S_I$ can be related to the charge carrier mobility $\mu$ and sheet carrier density $n$ respectively via

$$\mu = S_V \frac{L}{W} \tag{1}$$

and

**Table 1** The parameters of the CVD graphene based Hall sensor on foil before and after the encapsulation of PMMA.

|  | Before encapsulation | After encapsulation |
|---|---|---|
| Bias voltage (V) | 0.1 | 0.1 |
| Current ($\mu$A) | 83 | 53 |
| $\Delta V_H/\Delta B$ (V/T) | 0.0043 | 0.0064 |
| $S_V$ (V/VT) | 0.043 | 0.064 |
| $\mu$ (cm$^2$/V·s) | 1075 | 1600 |
| $S_I$ (V/AT) | 52 | 121 |
| $n$ (cm$^{-2}$) | 1.2×10$^{13}$ | 5.2×10$^{12}$ |

$$n = \frac{1}{eS_I}, \quad (1)$$

where e is the elementary charge [1,2]. The key parameters for the sensor before and after the encapsulation are summarized in Table 1. The current normalized sensitivity increases by 130%, and the corresponding decrease in sheet carrier density is also clearly shown, since the encapsulation isolates the graphene from ambient and hence reduces the doping from oxygen and water molecules in air. Meanwhile, the voltage normalized sensitivity increases by 50%, and the carrier mobility also increased correspondingly. Although the total resistance of the graphene increases as one can tell from the degraded current, the mobility still increased from 1075 cm$^2$/V·s to 1600 cm$^2$/V·s due to the encapsulation [12].

Two more sensor devices are characterized in a similar way and the sensitivities before and after the encapsulation are shown in Fig. 2. It is clearly demonstrated that the improvement in either voltage normalized or current normalized sensitivity after the encapsulation of PMMA is in general valid. The device to device variation of sensitivity values after the encapsulation should be noted, which can be attributed to the different thicknesses of the PMMA encapsulation layer from drop coating method and the relatively high water vapour transmission rate through PMMA. In this sense, the PMMA layer is not an appropriate encapsulation choice for a Hall sensor, but it is sufficient to prove the concept of the performance improvement. However, inhomogeneity of the CVD grown graphene also contributes to the device to device variation of the sensitivity.

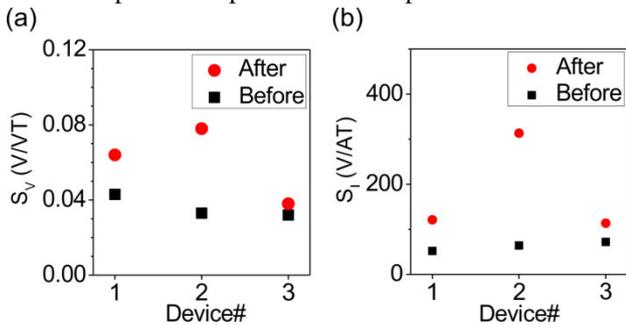

**Figure 2** The (a) voltage normalized and (b) current normalized sensitivity of three different Hall sensors based on large area CVD graphene on foil before and after the encapsulation with PMMA. The device shown in Fig. 1 as well as Table 1 is shown here as device number 1.

**4 Hall sensors encapsulated with hBN** In order to achieve higher sensitivity for the graphene based Hall sensor on foil, a proper encapsulation layer needs to be introduced. hBN is considered to be a perfect choice [13,14]. The homogenous van der Waals interaction between graphene and hBN guarantees very high carrier mobility as well as low carrier density of graphene, which is crucial for Hall sensor performance [5]. The illustration of the encapsulated Hall sensor is shown in Fig. 3a, and the optical image of the active area is shown in Fig. 3b, with inset showing the original hBN-graphene-hBN stack before patterning. The dimension of the graphene strip in the fabricated sensor has a length of 30 $\mu$m and a width of 12 $\mu$m. Due to the fabrication process, Hall measurements before encapsulation were not possible. For the Hall measurement after encapsulation, a bias voltage of 0.1 V was applied and the current was measured as 30 $\mu$A. The Hall measurement was carried out with magnetic fields of 6.0 mT, 12.0 mT, 18.0 mT, -18.0 mT, -12.0 mT, -6.0 mT in sequence, and the response over time is shown in Fig. 3c. It is also clear that the offset voltage of the Hall sensor is again very stable. Similarly the linear relationship between the Hall voltage and the applied external magnetic field is plotted in Fig. 3d, and the linear fitting of the data point gives a slope of 0.068 V/T, with a fitting error of $3.9 \times 10^{-4}$ V/T. The relative errors for the Hall voltage measurement are also shown for different magnetic fields, and all the errors are less than 0.7%. Based on the applied voltage bias and measured current, the voltage and current normalized sensitivity are 0.68 V/VT (with an error of $3.9 \times 10^{-3}$ V/VT from the fitting) and 2270 V/AT (with an error of 13 V/AT from the fitting) respectively, corresponding to a high charge carrier mobility of 17,000 $cm^2$/V·s including effects from the contact resistance and a low sheet carrier density of $2.8 \times 10^{11}$ $cm^{-2}$. In comparison with the four terminal carrier mobility obtained in Ref. [11], we assume that the mobility of the device here is underestimated with the presence of the higher contact resistance from the nickel contacts. The very high mobility is first thanks to the excellent quality of the CVD grown graphene, and the appropriate non-covalent encapsulation from hBN also preserves the intrinsically high mobility. Compared to the Hall sensor encapsulated with PMMA shown in Fig. 1, the carrier density here is decreased by more than one order of magnitude, resulting in a much higher current normalized sensitivity.

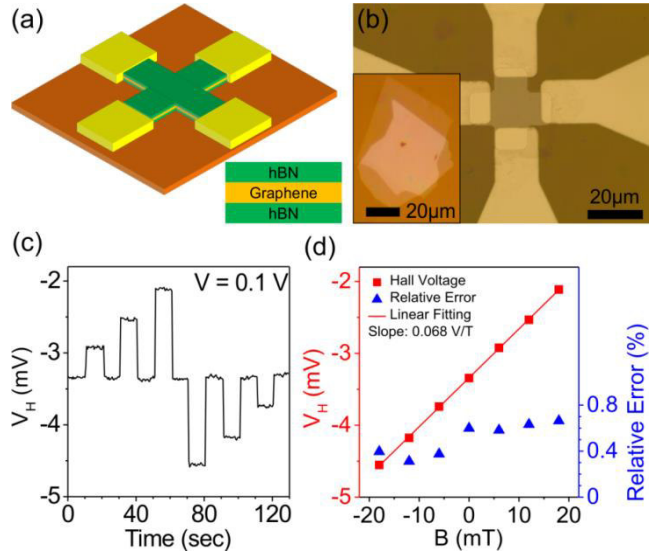

**Figure 3** The Hall sensor based on hBN encapsulated single crystal graphene on foil. (a) The illustration shows the geometry of the Hall sensor device. The encapsulated graphene is contacted at the edge after patterning. (b) The optical image of the active area of the Hall sensor with a scale bar of 20 $\mu$m. The inset is the original hBN-graphene-hBN stack before patterning, also with a scale bar of 20 $\mu$m. The length and width of the encapsulated graphene are 30 $\mu$m and 12 $\mu$m respectively. (c) The Hall voltage response to the applied magnetic field, with a sequence of 6.0 mT, 12.0 mT, 18.0 mT, -18.0 mT, -12.0 mT, -6.0 mT. Throughout the measurement, the bias is set to 0.1 V, and a current of 30 $\mu$A is measured. (d) The linear relationship between the Hall voltage and magnetic field, with a slope of 0.068 V/T from the linear fitting to the data points. The calculated voltage and current normalized sensitivity are 0.68 V/VT and 2270 V/AT respectively. The relative errors for the Hall voltage measurement are also shown.

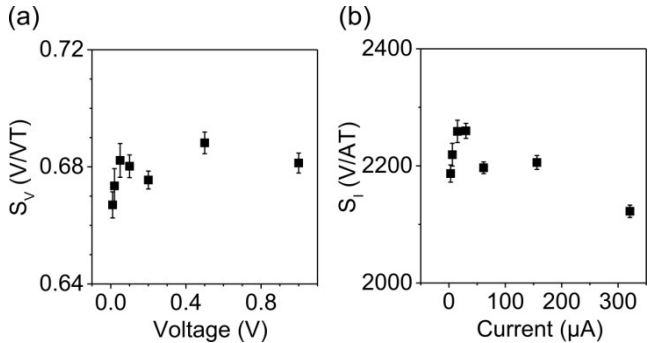

**Figure 4** The (a) voltage bias and (b) current bias dependence of the sensitivity of the Hall sensor based on hBN encapsulated graphene on foil. The error bars are shown for different biases.

The voltage and current bias dependence of the sensitivity values of the Hall sensor are tested and shown in Fig. 4. The voltage normalized sensitivity stays stable for voltage bias up to 1 V, which indicates that the mobility of the graphene is not notably affected by the Joule heating. On the other hand, the current normalized sensitivity shows no degradation up to 150 $\mu$A current bias. Afterwards, it starts to drop for a current bias of 300 $\mu$A, since the Joule heating from high current affects the doping level and hence the carrier density in graphene.

In Table 2, we listed the highest achieved sensitivity values from Hall sensors based on different materials and substrates. It is obvious that the graphene based Hall sensors are able to achieve higher sensitivities in general if an encapsulation layer is applied. Among all the Hall sensors on foil, the one with hBN encapsulation in this work shows the highest sensitivity so far, which surpasses the performance of silicon based Hall sensor by a factor of 22.7 for current normalized sensitivity and by a factor of 6.8 for voltage normalized sensitivity.

**Table 2** The comparison of the sensitivities between Hall sensors with different materials and substrates.

|  | Substrate | Encapsulation | $S_I$ (V/AT) | $S_V$ (V/VT) |
|---|---|---|---|---|
| CVD graphene (This work) | Foil | hBN | 2270 | 0.68 |
| CVD graphene (This work) | Foil | PMMA | 313 | 0.078 |
| CVD graphene with SAM [6] | Foil | - | 437 | 0.134 |
| Bismuth [15] | Foil | - | 2.3 | - |
| Si [16,17] | Rigid | - | 100 | 0.1 |
| Exfoliated graphene [5] | Rigid | hBN | 5700 | 3 |
| CVD graphene with SAM [6] | Rigid | - | 2364 | - |

**5 Conclusion** In summary, the encapsulation of the graphene based Hall sensors on foil is performed and is able to provide higher sensitivity for magnetic field detection. After a simple PMMA encapsulation, improvement in both current and voltage normalized sensitivity can be obviously observed. Furthermore, when hBN encapsulation is applied, the current and voltage normalized sensitivities reach 2270 V/AT and 0.68 V/VT respectively, which are the highest sensitivity values for Hall sensors on foil. Therefore the encapsulation can be considered as a very effective strategy to improve the performance in terms of higher sensitivity of a Hall sensor on foil.

**Acknowledgements** This work was financially supported by the European Commission under the projects Graphene Flagship (contract no. 604391), and by the German Science Foundation (DFG) within the priority program 1796 FFlexCom Project "GLECS" (contract no. NE1633/3).